\documentclass{ws-ijmpd}
\usepackage[super,compress]{cite}
\usepackage{amsmath,overpic}
\usepackage[colorlinks=true,allcolors=blue]{hyperref}
\begin{document}

\markboth{GIUSEPPE CONGEDO}
{DETECTION PRINCIPLE OF GRAVITATIONAL WAVE DETECTORS}

%
\catchline{}{}{}{}{}
%

\title{DETECTION PRINCIPLE OF GRAVITATIONAL WAVE DETECTORS}

\author{GIUSEPPE CONGEDO}

\address{Institute for Astronomy, School of Physics and Astronomy, University of Edinburgh, \\
Royal Observatory, Blackford Hill, Edinburgh, EH9 3HJ, United Kingdom \\
\vspace{1pt}
Department of Physics, University of Oxford, \\
Keble Road, Oxford OX1 3RH, United Kingdom \\
\vspace{1pt}
giuseppe.congedo@ed.ac.uk}

\maketitle

\begin{history}
\received{Day Month Year}
\revised{Day Month Year}
\end{history}

\begin{abstract}
With the first two detections in late 2015, astrophysics has officially entered into the new era of gravitational wave observations. Since then, much has been going on in the field with a lot of work focussing on the observations and implications for astrophysics and tests of general relativity in the strong regime. However much less is understood about how gravitational detectors really work at their fundamental level. For decades, the response to incoming signals has been customarily calculated using the very same physical principle, which has proved so successful in the first detections. In this paper we review the physical principle that is behind such a detection at the very fundamental level, and we try to highlight the peculiar subtleties that make it so hard in practice. We will then mention how detectors are built starting from this fundamental measurement element.
\end{abstract}


\ccode{PACS numbers: 04.80.Nn; 04.30.-w; 04.20.-q.}


\section{Introduction}	

Gravitational wave (GW) detectors have now officially entered into the era of GW astronomy with the first ever observation of two signals from black hole binaries in late 2015 \cite{abbott2016a, abbott2016b}. These events alone have already produced stringent tests of general relativity in the strong regime \cite{abbott2016c}, and constraints on the source distributions and evolution models for these unknown types of astrophysical sources \cite{abbott2016d}.

However, how GW detectors really work in detecting these waves remains a bit obscure. We are so accustomed to observing the universe through the whole electromagnetic spectrum, that observing it with GWs becomes now less obvious. One striking difference is that we will never be able to produce direct images of GW binaries. Instead, what we can certainly do is to infer their properties by looking at how the detector responds to the incoming signal. Early work to answer this question focussed on resonant bar detectors, and the response was described in terms of geodesic deviation and Riemann tensor \cite{weber1960,isaacson1968}. However, the approach of using the metric perturbations took over in the 70s with the first ideas about interferometers \cite{forward1978}, and also with spacecraft tracking experiments \cite{estabrook1975}. The main problem with it lies in the misusage of the metric perturbations to derive physical interpretations that are often dependent upon the choice of the gauge transformation \cite{koop2014}. Evidently this is not the case of the early attempts with the Riemann tensor, which directly translates into relative acceleration between test masses, and therefore behaves as a proper physically observable quantity.

\begin{figure}[!htbp]
\centering
\begin{overpic}[width=.7\textwidth]{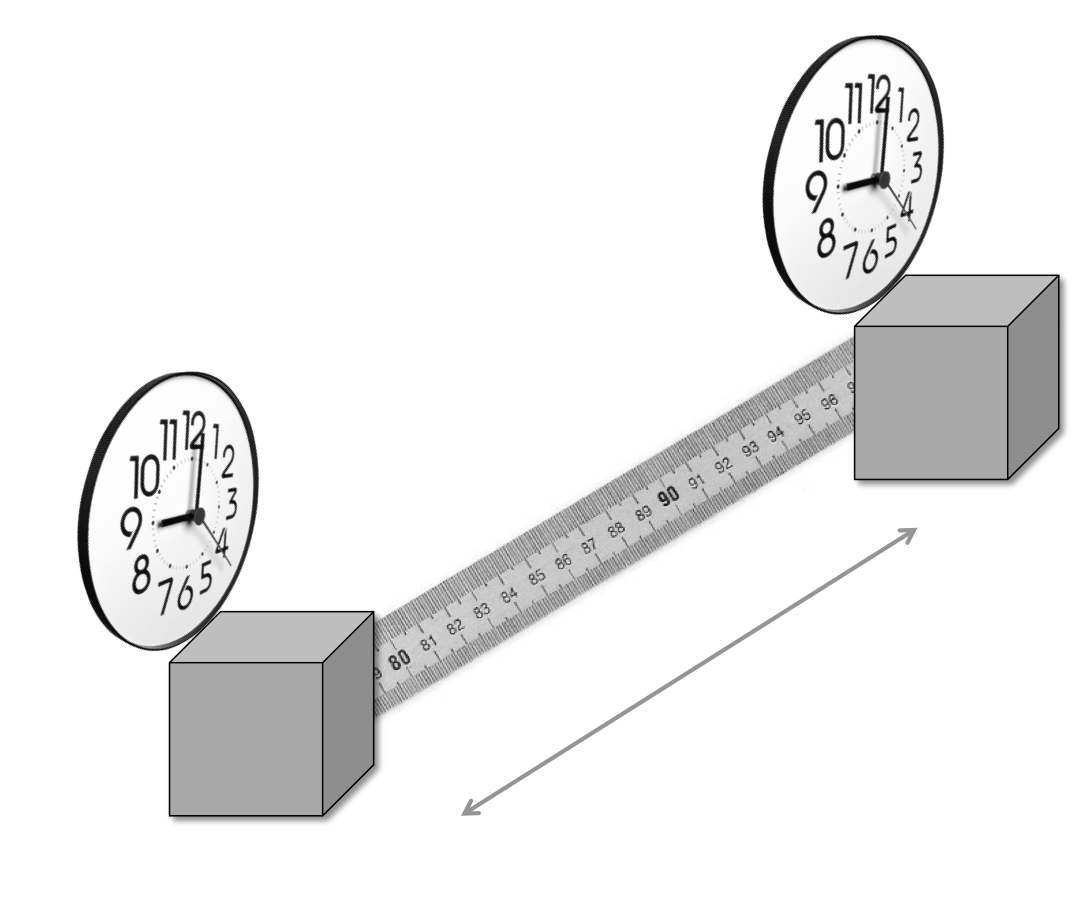}
\put (75,20) {$\delta a / L\sim\ddot{h}$}
\end{overpic}
\caption{\footnotesize{Simple idealised mockup of the detection principle. A single arm gravitational wave detector -- the measurement element -- is composed by three key parts: two free falling test masses, two accurate clocks, and an accurate \textit{spacetime metre stick}. The fidelity of free fall and the perfectness of clocks and metres are generally the limits for any gravitational wave detection. The response of the system can be described in terms of relative acceleration in a fully covariant way -- see text for details -- and the observed relative acceleration, $\delta a$, is proportional to the second derivative of the metric perturbation, $h$.}}
\label{fig:detection_principle_mockup}
\end{figure}

In fact little known is that any modern interferometric detector can, in principle, be described in terms of the Riemann tensor acting on the fundamental measurement element (see Fig.\;\ref{fig:detection_principle_mockup}) \cite{congedo2015b,koop2014,congedo2013}. This is composed by only three constituents: two free falling test masses, two accurate clocks, and an accurate \textit{spacetime metre stick}. As simplified this could be, it would be already enough, with infinite accuracy, to respond to a GW by producing relative acceleration between the test masses, and record this motion.

\begin{figure}[!htbp]
\centering
\begin{overpic}[width=.7\columnwidth]{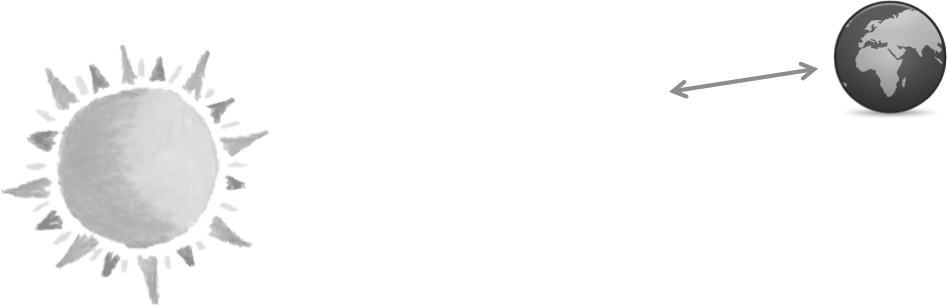}
\put (65,21) {$L_1$}
\put (75,10) {$\delta a/a= \eta\,\Delta\Omega$}
\end{overpic}
\caption{\footnotesize{Not only is the relative acceleration between test masses a convenient quantity to describe the effect of gravitational waves on detectors, but also the physical observable in other experiments too, including tests of the strong equivalence principle with the Sun-Earth collinear Lagrangian points \cite{congedo2016}, where planets act as test masses and their positions are tracked with radio signals. Note that the observed relative acceleration is proportional to the Nordtvedt parameter, $\eta$, and the difference between self-energies, $\Delta\Omega$.}}
\label{fig:eta_detection_mockup}
\end{figure}

The concept of relative acceleration is so general that it is a useful quantity to characterise the physical observable in other experiments too. For instance, this is the case for tests of the strong equivalence principle with the Sun-Earth collinear Lagrangian points \cite{congedo2016} (see Fig.\;\ref{fig:eta_detection_mockup}). In effect, relative acceleration is at the heart of gravitation and general relativity, in a way that it becomes particularly important in all non-local calculations.

In this paper we will give a flavour of what a detection really is, from the generated wave in the source's reference frame (Sec.\;\ref{sec:signals}), to the traditional way of deriving the detector's response (Sec\;\ref{sec:meas_principle_traditional}). We will go through the more recent revisitation \cite{congedo2015b} of the traditional frequency shift calculation (Sec.\;\ref{sec:meas_principle_recent}) that clarifies how the detector responds to GWs, including other effects that act as nuisance for a detection. Compared to previous work \cite{koop2014,congedo2013} and more traditional work, this is done in a fully covariant, gauge and coordinate independent way. The main advantage is that the frequency shift can be derived directly from first principles in general relativity. We will then mention how measurement elements are combined in different detectors, both on ground and in space (Sec.\;\ref{sec:detectors}).

\section{Gravitational wave signals} \label{sec:signals}

We briefly review how GWs are produced up to the first parametrised post-Newtonian (PPN) term. More accurate templates are of course calculated using higher order corrections and numerical relativity. The interested reader is invited to take a look at Ref.\;\refcite{maggiore} for a more thorough discussion.

Pictorially, these waves are described as ripples in the spacetime fabric -- in practice they are first order perturbations of an underlying metric, which we assume flat. Mathematically, $g_{\mu\nu}=\eta_{\mu}+h_{\mu\nu}+\mathcal{O}(h^2)$, where the perturbation evolves with time. The evolution with time is given by the Einstein equations. To work it out, one needs to calculate the Riemann tensor up to $\mathcal{O}(h^2)$ and finally get to the following wave equation
\begin{equation}
\bar{h}^{~~~\sigma}_{\mu\nu,~\sigma}=
\begin{cases} 
\frac{16\pi G}{c^4}T_{\mu\nu} & \text{inside the source} \\
0 &  \text{outside the source}
\end{cases}.
\end{equation}
It is worth noting that while the wave equation is gauge dependent -- note that $\bar{h}_{\mu\nu}$ is written in the traceless-transverse (TT) gauge -- any equation involving directly the Riemann tensor, by construction, will automatically cancel out all the additional gauge-dependent terms appearing at the level of the metric tensor. We will use this fact later on in this paper to introduce a gauge independent formalism for the detector response to GWs. Also please note that the physical significance of the TT gauge is that, under the assumption of a weak wave, the propagation is \textit{transverse} (effects on test masses are orthogonal to the direction of propagation) and \textit{traceless} (there is no gravitational source term pumping up the wave).

Meanwhile let us take a look at how the signal is generated by a merging binary. A so-called chirping binary is characterised by three phases:
\begin{enumerate}
\item \textit{Inspiral}: a slow increase in frequency/amplitude, fully described by analytical formulae to the lowest PPN order -- the two objects approach each other.
\item \textit{Merger}: a quick increase in frequency/amplitude in the last few cycles or so, before the actual merge, which can be described by higher PPN terms or numerical relativity -- the two objects start merging each other.
\item \textit{Ringdown}: an exponential decay in frequency/amplitude, where the merged object is fully described, again, by analytical formulae -- the final object loses gravitational energy and moment of inertia until no further radiation is emitted.
\end{enumerate}

It is important to distinguish between observed quantities and the same evaluated at the source's reference frame, also called \textit{rest frame}. The relation between the two is pretty straightforward and it involves the source redshift. Therefore the observed proper time, frequency, and mass of the binary are given by
\begin{align}
\text{d}t'&=(1+z)\text{d}t, \\
f'&=(1+z)^{-1}f, \\
M'&=(1+z)M,
\end{align}
where the primed quantities are observed, and the unprimed ones are in the rest frame.
These relations become important when dealing with sources at cosmological distances, but even for the first ever detected source the correction is just $\sim10\%$ (as $z\sim0.1$, assuming a $\Lambda$CDM cosmology). The other fact is that any detector measures $M'$, and therefore a direct measurement of $M$ is impossible without assuming or independently measuring the redshift.

The evolution of the GW signal is determined primarily by the its phase, $\Phi(t)$, which is related to the frequency, $f(t)$, by $\Phi(t)=2\pi\int f(t')\text{d}t'$. The frequency is given by solving the dynamical equation of motion of the binary, which can be quite complex depending on the PPN order or accuracy of the full numerical relativity calculation. However the differential equation that relates $\dot{f}$ to $f$ can be very simple for an inspiral at the lowest PPN term,
\begin{equation}
\frac{\text{d}f}{\text{d}t}=\frac{96}{5}\,\pi^\frac{8}{3}\left(\frac{GM}{c^3}\right)^\frac{5}{3}f^\frac{11}{3},
\end{equation}
where $M=(m_1 m_2)^{3/5}/(m_1 + m_2)^{1/5}$ is the so-called \textit{chirp mass} that determines the frequency evolution of the system at the same PPN term.

In the wave's coordinate system (where the wave propagates along the $z$ axis) and in the TT gauge, the two metric components, i.e.\ the GW signal, are
\begin{align}
h_+(t)&=A(t)\frac{1+\cos^2\iota}{2}\cos\Phi(t),\\
h_\times(t)&=A(t)\cos\iota\sin\Phi(t),
\end{align}
where $\iota$ is the inclination angle: $\iota=0$ for edge-on sources and therefore circularly polarised signals; $\iota=\pi/2$ for face-on sources and therefore linearly polarised signals. Also $A(t)$ is the signal amplitude that is inversely proportional to the luminosity distance $d_L$. Inspiral, merge, and ringdown are characterised by different evolutions for $A(t)$ and $f(t)$. At the lowest PPN order, the same equations hold true for inspiral and merge (although less accurately for the latter), and thus
\begin{equation}
A(t)=
\begin{cases}
\frac{4}{d_L(z)}(\frac{GM(z)}{c^2})^{5/3}(\frac{\pi f(t)}{c})^{2/3} & \text{inspiral+merge} \\
\frac{4\pi^2 G}{c^4}\frac{I_3 f^2(t)}{d_L(z)}\epsilon & \text{ringdown}
\end{cases},
\end{equation}
where we have denoted $M(z)=(1+z)M$; $I_3$ and $\epsilon$ are, respectively, the moment of inertia and ellipticity of the oblate merged object.

Given the frequency evolution of the system, this model is already enough to provide the signal template of a GW source just before detection -- we are still missing how the signal affects the detector's dynamics.  

\section{Measurement principle: the traditional approach} \label{sec:meas_principle_traditional}

The traditional approach when calculating the response of detectors to GWs can be summarised in two alternative methods:
\begin{enumerate}
\item \textit{Light travel}. The response is calculated by integrating the null geodesic equation,
\begin{equation}
\text{d}s^2=0,
\end{equation}
of light bouncing off free falling test masses. The effect of GWs is such that the light travel time is changed with respect to the nominal delay corresponding to the distance between the test masses. Therefore the observer detects light phase differences. \\
\item \textit{Geodesic deviation}. The response is calculated by integrating the geodesic deviation,
\begin{equation}
\frac{\text{D}^2{\delta x^\mu}}{d\tau^2}=R^\mu_{~\alpha\beta\gamma}v^\alpha v^\beta\delta x^\gamma,
\end{equation}
between two neighbouring world lines with 4-velocity $v^\mu$ and separated by $\delta x^\mu$. The effect of GWs is such that a relative acceleration is induced in addition to the Newtonian acceleration. Therefore the observer detects differential forces.  
\end{enumerate}
It is worth noting that the first method is the standard with interferometric detectors and pulsar timing arrays, whereas the second was traditionally used in the past with bar detectors.

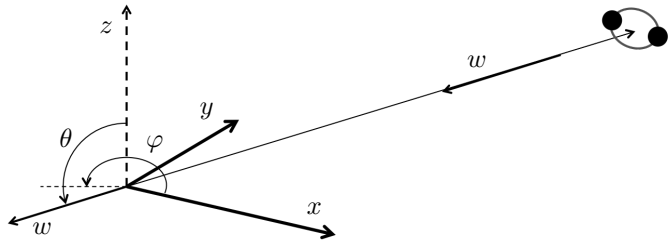
\begin{figure}[!htbp]
\centering
\begin{overpic}[width=.7\columnwidth]{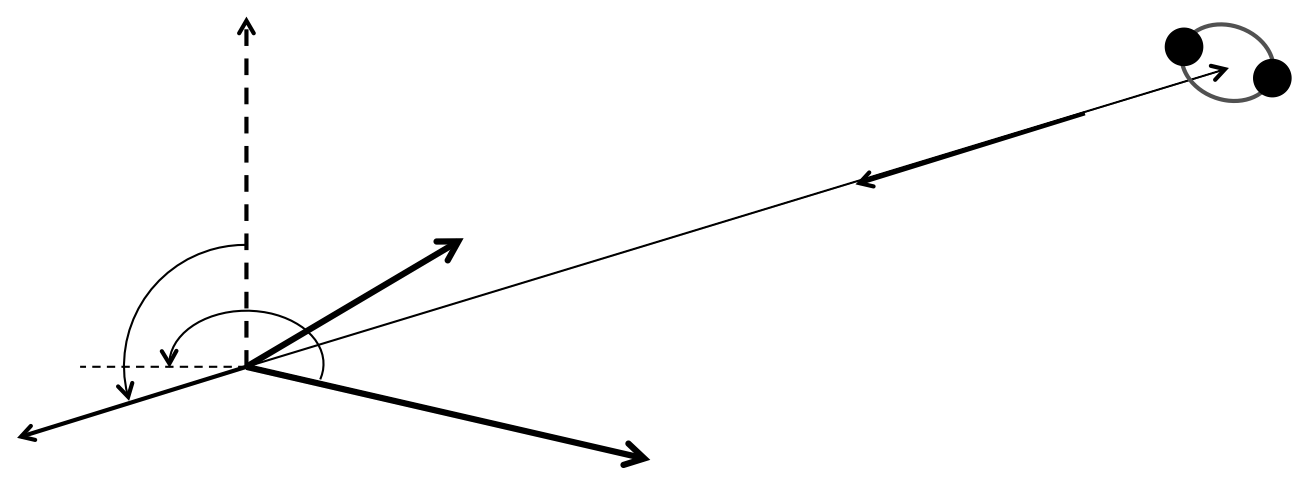}
\put (70,27) {$w$}
\put (5,2) {$w$}
\put (9,15) {$\theta$}
\put (22,15) {$\varphi$}
\put (46,5) {$x$}
\put (30,20) {$y$}
\put (15,32) {$z$}
\end{overpic}
\caption{\footnotesize{Schematic diagram of a Michelson detector and its coordinate system (axes in bold). The incoming wave is described by the vector $w$, with colatitude $\theta$ and longitude $\varphi$. The response to the incoming GW is worked out in this coordinate system -- see text.}}
\label{fig:michelson_response}
\end{figure}

As an illustrative example, let us consider an equal arm Michelson interferometer with armlength $\delta x$. One can define a coordinate system (see Fig.\;\ref{fig:michelson_response}) that is fixed to the detector itself, with the $x$ axis along one arm, $y$ along to the other arm, and $z$ orthogonal to the detector's plane. In the long wavelength limit \footnote{When the GW frequency does not change significantly during the light travel and therefore $\delta x\ll\lambda$, which is relevant for ground-based, but not for space-based detectors where the light travel delay must be correctly taken care of.}, the interferometric response is described in terms of phase differences
\begin{equation}
\delta\phi(t)=\frac{4\pi\delta x}{\lambda_\text{laser}}h(t),
\end{equation}
where $\lambda_\text{laser}$ is the light wavelength, and $h$ is the strain, which is given by a linear combination of the wave components
\begin{equation}
h(t) = F_+(\theta,\varphi)h_+(t)+F_\times(\theta,\varphi)h_\times(t),
\end{equation}
and $F_+$ and $F_\times$ are the antenna response functions, both implicitly dependent on time if the source's angular position with respect to the detector, i.e.\ the angles $\theta$ and $\varphi$, depends on time too. Explicitly,
\begin{align}
F_+(\theta,\varphi)&=\frac{1+\cos^2\theta}{2}\cos{2\varphi}, \\
F_\times(\theta,\varphi)&=-\cos\theta\sin{2\varphi}.
\end{align}
We can now see that the effect of an incoming GW is to produce a phase difference at the output of the detector, $\delta\phi$. In the long wavelength approximation, this phase difference is just a linear combination of the two metric components in the TT gauge, which corresponds to a rotation from the wave's coordinate system to the detector's coordinate system. If the position of the source with respect to the detector changes with time (because the detector itself is moving around), then the coefficients are time dependent. This time dependence is efficiently used for sky localisation. In fact, as the detector moves around, a frequency/amplitude modulation of the incoming wave is induced in the detector output. If the source is itself modulated (e.g.\ because the binary is inspiralling), the combination of detector and source modulation allows a much better sky localisation. Also, we can see that $F_+=0$ for $\varphi=\pm\,\pi/4$ and $F_\times=0$ for $\varphi=\pm\,\pi/2$ or $\theta=\pm\,\pi/2$, therefore the detector is blind to either of the two polarisations, with worse sky localisation. Clearly, sky localisation for a given detector is a direct function of polarisation through angular position and detected signal to noise.

Similarly, if one cannot work in the long wavelength approximation, and wants to allow for the light travel delays to be fully taken into account, the response of an unequal-arm Michelson interferometer is best described in terms of fractional frequency shifts along individual arms \cite{vinet2013}. These are linearly combined, with proper time delays accounting for the round trip of the photon along the arms, to cancel the laser frequency noise  exactly \cite{tinto2014}, which is nil by construction in an equal-arm interferometer. This complication becomes strictly necessary for space-based detectors, whereas the equations introduced above would suffice for ground-based detectors, but similar arguments hold true for sky localisation vs polarisation sensitivity. We omit the relevant formulae here for brevity.

One drawback of the whole picture presented so far is that it lacks an explicit description of the main measurement limitations: (\textit{i}) geodesics are not real geodesics -- there are non-gravitational forces that perturb the test masses away from their free fall; (\textit{ii}) geodesics are never infinitely close to each other; (\textit{iii}) detector noise always affects the measurement, e.g.\ interferometry, thermal noise in bar detectors, or timing jitter in pulsar timing array; (\textit{iv}) as test masses are not in free fall, they must also be subjected to inertial forces as well.

\section{Measurement principle: a more recent perspective} \label{sec:meas_principle_recent}

As already seen in the previous section, there are two approaches one may take when working out the detector's response to GWs. The first relies on the metric to calculate the phase shift induced by a change of the optical path; instead, the second is based on the Riemann tensor to derive the relative acceleration induced by a change of the underlying metric. Of course, they are radically different, but somehow related one to the other. In this section we will review the equivalence between the two in a more unified way.

The geodesic deviation approach has been predominantly used during early attempts with resonant bars, until the interferometric approach with the metric perturbations took over in the 70s. Because interferometers employ rate of change in the optical path to quantify the effect of incoming GWs, it does make sense to use the light travel approach, thus the metric itself. However, using the metric perturbation may be very dangerous. As this quantity is gauge dependent, assuming a gauge in the first place might take to ambiguous gauge-dependent results. Evidently the metric itself is not a proper physical observable of the system. This has led to a variety of somewhat coloured descriptions of how a detector responds to a GW, as cleverly pointed out in Ref.\;\refcite{koop2014}, where we defer the interested reader to. All of these descriptions are obviously wrong, but of course they hide a partial truth underneath.

Recent attempts to address this problem tried to revisit the traditional calculation of the light frequency shift with alternative approaches \cite{congedo2013,koop2014}. Now geodesics are more physical than ever, with non-gravitational forces acting upon them, and light rays bouncing on and off tests masses in quasi free fall. Yet again, although these approaches show very similar results, there is a substantial difference in that one method solves the problem by using a non-standard time-like congruence between geodesics, and the other one employs a null congruence between present and past geodesics. An even more recent paper \cite{congedo2015b} showed that the frequency shift can be directly related to the Riemann tensor -- the only meaningful physical observable -- in a fully covariant and gauge independent way that does not need any formulation of a congruence at all. Instead, everything can be derived from first principles, such as the parallel transport of 4-vectors. That seems to be enough to relate the frequency shift to the Riemann tensor. However, in doing so, other effects pop out, whose physical interpretation will now become clear.

\begin{figure}[h!]
\centering
\begin{overpic}[width=0.7\textwidth]{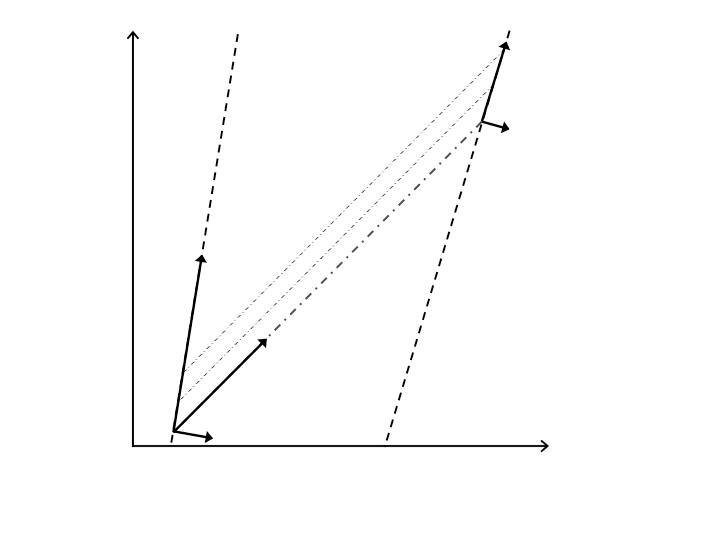}
\put (14,66) {$ct$}
\put (73,9) {$x$}
\put (22,29) {$u^\mu$}
\put (70,62) {$v^\mu$}
\put (33,21) {$k^\mu$}
\put (21,15) {e}
\put (64,58) {r}
\put (67,53) {$g^\mu$}
\put (30,15) {$f^\mu$}
\put (29,38) {\rotatebox{81}{emitter's geodesic}}
\put (56,17) {\rotatebox{74}{receiver's geodesic}}
\put (34,29) {\rotatebox{45}{light beam's geodesic}}
\end{overpic}
\vspace{-20pt}
\caption{\footnotesize{Adapted from Ref.\;\protect\refcite{congedo2015b}. Instantaneous Minkowski diagram for the thought experiment of two test masses exchanging light rays and measuring the corresponding frequency shift. Null geodesics connect the emission event ``e'' to the reception event ``r''. These intersect the emitter and receiver's geodesics at those specific events in spacetime. 4-velocities, 4-forces, and the light beam's 4-momentum are also shown. A variation of the fractional frequency shift between ``e'' and ``r'' is induced by a change in the Riemann tensor, which arises from a parallel transport of vectors between those events. Other contributions also affect the measurement, including fictitious forces.}}
\label{fig:diagram}
\end{figure}
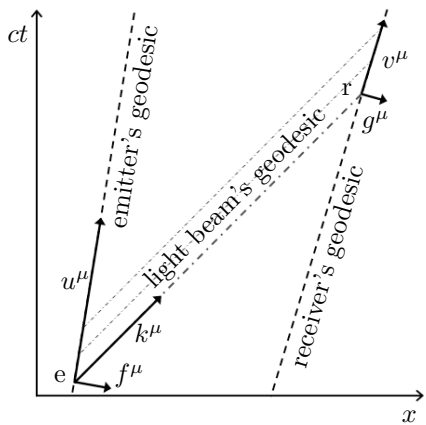

Let us go through the thought experiment already presented in Ref.\;\refcite{congedo2015b} and also shown in Fig.\;\ref{fig:diagram}. The key measurement element of a GW detector is based on two test masses exchanging light rays. This is true for all interferometric detectors, both on ground and in space, and also for pulsar timing array (except that the observable is the pulse timing instead of the frequency shift), but it is not strictly applicable to bar detectors \footnote{Although an alternative formulation might be derived and adapted to this case too.}. At a given time, a test mass is treated as the emitter of a single light pulse, and the other one as the receiver of the same pulse at a later time. Therefore two events, emission ``e'' and reception ``r'', live separately in the spacetime, and they are casually connected through the light's null geodesic. In fact, the emitter's geodesic, with 4-velocity $u^\mu$, intersects the null geodesic at ``e'', and the receiver's geodesic, with 4-velocity $v^\mu$, intersects it at ``r''. Also, both test masses are not in free fall, i.e.\ they are subjected to non-gravitational forces (per unit mass) $f^\mu$ and $g^\mu$. If $k^\mu$ is the light 4-momentum, the receiver will measure the light frequency $\omega_\text{r}=k_\mu(\text{r})v^\mu(\text{r})$, whereas the emitter will measure $\omega_\text{e}=k_\mu(\text{e})u^\mu(\text{e})$. Therefore their difference returns the frequency shift that should be sensitive to gravitational effects.

The full calculation is described in length in Ref.\;\refcite{congedo2015b}, but here we report on the main result. If one works in the reference frame of the receiver -- as it should do as measurements are made in this reference frame -- and parallel transports all quantities from ``e'' to ``r'', the first derivative of the frequency shift with respect to the proper time of the receiver, $\tau_\text{r}$, yields 
\begin{equation}
\label{eq:freq_diff_der_final}
\frac{\text{d}\delta\omega}{\text{d}\tau_\text{r}}=
k_\mu(\text{r})
\mathcal{R}^\mu
+\frac{\text{D} k_\mu}{\text{d}\tau_\text{r}} \left[v^\mu(\text{r})-u^\mu(\text{e})\right]
+k_\mu(\text{r})\left[g^\mu(\text{r})-f^\mu(\text{e})\right]
+\gamma_\text{fict}\left(\Gamma\right).
\end{equation}
Here $\mathcal{R}^\mu$ is the Riemann tensor, contracted with both 4-velocities and the light's 4-momentum, integrated over the light path from ``e'' to ``r''. This is the main result: the first time-derivative of the frequency shift gives an integrated measure of the Riemann tensor through a non-local integral. Of course, this term recovers to standard results when we consider the case of two neighbouring geodesics, low velocities, and the calculation is done in the local Lorentz frame. But what should also strike our attention is the additional terms:
\begin{enumerate}
\item \textit{Doppler term due to the rotation of the line of sight}: this is caused by the apparent motion of the emitter in the reference frame of the receiver.
\item \textit{Differential non-gravitational forces}: these are the cause of the non-free-fall motion of both test masses.
\item \textit{Inertial forces}: these depend on all $\Gamma$s and therefore vanish in the local Lorentz frame.
\end{enumerate}
This fully covariant and gauge free framework shows what the measurement element of GW detectors is sensitive to. Along with GWs obviously entering into the Riemann tensor, other effects are also picked up -- notably, non-gravitational forces and inertial forces.

It is worth noting that, in order to derive the above result, we have employed only first principles, with no further approximations. It should now be clear that the effect of an incoming GW on the measurement element -- at the fundamental level -- is to translate a variation of the Riemann tensor into a net displacement of vectors when parallel transported along the null geodesic connecting emission to reception, and the observable effect is always a relative acceleration.

\section{Detectors as combinations of the fundamental measurement element} \label{sec:detectors}

We will now focus our attention on real detectors, and mention how they can be constructed as clever combinations of the fundamental measurement elements. For instance, the equal arm Michelson interferometer introduced in Sec.\;\ref{sec:meas_principle_traditional} comprises two measurement elements. The laser light is first split, and then recombined at the output to make interferometry and, by construction, this configuration suppresses the laser frequency noise.

The first example is a ground-based detector, like LIGO that detected GW signals from binary black holes inspirals\cite{abbott2016a,abbott2016b}. It is a Michelson interferometer that works in the $10\;\text{Hz} - 1\;\text{kHz}$ band with typical armlength of $3-4$\;km. Here Fabry-Perot resonant cavities improve the laser power such that the shot noise -- the main noise source at mid-high frequency -- is reduced by two orders of magnitude, down to $10^{-23}\;\text{Hz}^{-1/2}$ in strain sensitivity. Other noise sources are: thermal noise, due to the fact that big mirrors acting as free-falling test masses are suspended with fused silica; and, above all, seismic noise that limits all measurements below 10\;Hz, and is mitigated with high quality factor multi-stage pendulums, which the mirrors are suspended to. The outputs of more detectors are combined at the level of measurement likelihoods to further mitigate noise sources, and help discriminate real signals from spurious signals. The current network of advanced detectors include the two LIGOs and VIRGO. The future of ground-based detectors lies in the ability to improve the hardware, and efficiently develop a bigger network of detectors. One key example is the Einstein Telescope \cite{hild2012} where two types of interferometers, making a total of six, are combined in a triangle shaped underground configuration of size 100\;km. The high frequency detectors would use the same technology employed in the current detectors, whereas the low frequency detectors would use cryogenic, low power, squeezed light, thus extending the frequency band down to 1\;Hz. Combining multiple detectors, better of different types, ultimately improves sensitivity and sky localisation, especially at low frequency.

The same principle of combining multiple measurement elements is also true for the planned space-based detector LISA \cite{elisa2013}, whose key technologies and instrument noise have been successfully tested with LISA Pathfinder \cite{armano2016,congedo2015a}. In this case the frequency band is much lower, $0.1 - 100$\;mHz, with a much bigger armlength, $\sim10^6$\;km. A simple argument for going into space is the following. As $M\sim f^{-8/5}$, and $d_L\sim f^{-2}$, at lower frequencies the detector becomes more sensitive to bigger masses and bigger luminosity distances, impossible to see on ground as seismic noise limits all measurements below 1\;Hz.   Typical detectable sources would be merging supermassive black hole binaries at cosmological distances, galactic binaries (either monochromatic or chirping), and extreme mass ratio inspirals. The instrumental noise sources of such a detector \footnote{Bearing in mind that a foreground of unresolved galactic binaries will be effectively treated as an additional noise source at low frequency.} are completely different from ground experiments: force noise limits all measurements especially at low frequency, mostly coming from Brownian noise and electrostatic actuation; interferometry and shot noise become more important at mid-high frequency. The design of LISA is such that three linearly dependent unequal arm interferometers, in orbit around the Sun, are combined in a way that the strain sensitivity would be at around $10^{-21}\;\text{Hz}^{-1/2}$ or better, and a sky localisation of $1\;\text{deg}^2$ or less. All the six individual outputs of the measurement elements are time shifted and linearly combined to derive the three synthetic interferometers that beat down the laser frequency noise -- this makes the LISA detector.

Much beyond in the future, and yet to be an approved mission, is BBO \cite{cutler2009}, a network of GW detectors in space that would combine more LISA-like interferometers, with shorter armlengths. The focus would be on GW cosmology, by measuring cosmological parameters with less than a percent accuracy. One master detector would comprise two overlapping LISA-like detectors -- this would give better strain sensitivity, which becomes critical for primordial background detection. Further two slave detectors in opposition would give a longer baseline, therefore a better angular resolution necessary for foreground subtraction.

All these examples of detectors, from ground to space, show that the measurement element of GW detectors is always the same -- two test masses and a laser beam -- and yet the way it is combined in a detector, or a network of detectors, can address very different GW science with very different measurement problems.

\section{Conclusions}

In this paper we have briefly reviewed the essence of GW detection at a very fundamental level, emphasising how the measurement element -- comprising two test masses and a laser beam -- is sensitive to GWs. This measurement element responds to fluctuations of the Riemann tensor integrated over the light path from emission to reception, which is a somewhat different approach compared to traditional calculations. The main advantage is the covariant gauge-free formulation, which eases the physical interpretation. The physical observable becomes the rate of change in the frequency shift, which is directly related to relative acceleration. However, this quantity also picks up other nuisance, e.g.\ non-gravitational forces that act on the test masses and disturb their free fall, and inertial forces that appear as observations are not made in an inertial reference frame. It is worth noting, though, that inertial forces do not play a significant role in ground-based detectors as the frequencies at which this affect becomes relevant are much lower than the measurement band. A different situation would be for space-based detectors as the effect falls in band -- this should be taken into account and ultimately corrected for in the calculation of the detector's response. We have reviewed, far from being exhaustive, how detectors are built as combinations of the fundamental measurement element. We have mentioned their main differences and peculiarities, how different combinations address different science questions, and yet the fundamental measurement principle is ultimately the same. 

\section*{Acknowledgments}

The author acknowledges support from Hertford College, Harding Fund, 
the Beecroft Institute for Particle Astrophysics and Cosmology, and Oxford Martin School.  



\bibliographystyle{ws-ijmpd}
\bibliography{references}

\end{document}